\documentstyle[prl,aps,multicol]{revtex}

\newcommand{\be}{\begin{eqnarray}}
\newcommand{\ee}{\end{eqnarray}}
\newcommand{\non}{\nonumber\\}
\newcommand{\inline}[1]{\noalign{\hbox{#1}}}
\newcommand{\ave}[1]{\left\langle #1 \right\rangle}


\begin{document}

\title{Fluctuations of particle ratios and the abundance of hadronic
  resonances}

\author{S. Jeon\thanks{
e-mail: sjeon@lbl.gov
} and V. Koch\thanks{
e-mail: vkoch@lbl.gov
}}

\address{Nuclear Science Division\\
Lawrence Berkeley National Laboratory\\
Berkeley, CA 94720, USA}

\maketitle

\begin{abstract}
In this letter we will argue that the event-by-event fluctuations of the ratio
of positively over negatively charged
pions provides a measurements of the number
of rho and omega mesons right after hadronization. This finding can be
utilized to put the hypothesis of chemical equilibration in relativistic heavy
ion collisions to a test.
\end{abstract}

\begin{multicols}{2}
The question to which extent the matter created in relativistic heavy ion
collisions is equilibrated is central to the interpretation of many
observables for the existence of a new phase of matter. 
A detailed analysis of the inclusive single particle yields of several
hadronic species has led many authors 
\cite{BGS98,BHS99,CR99,Hei99} 
to conjecture that rather early in the collision
chemical equilibrium has been reached. Indeed assuming chemical equilibrium at
an early stage of the collisions a 
rather impressive agreement with a large body of data can be obtained by
adjusting just a few parameters, namely the temperature, the baryon chemical
potential and the strangeness suppression factor (for details see e.g.
\cite{BHS99}). However, this analysis has to rely on the abundance of final
state `stable' particles and thus has to infer the number of most 
hadronic resonances present inside the system. Some information about
the abundance of unstable resonances, such as the $\rho$ and $\omega$ meson,
may be obtained through the observation of the electromagnetic decay into
dileptons. However, this only provides a time integrated yield and thus gives
only limited information about the abundance of these resonances right after
hadronization and/or chemical freeze-out. 

In this letter we propose to study the event-by-event  
fluctuations of particle ratios, in
particular the ratio $\pi^+/\pi^-$
in order to put a strong constraint on the relative abundance of some
unstable resonance right after hadronization/chemical freeze-out. We will show
that the fluctuations of $\pi^+/\pi^-$ are quite sensitive to the number of
the {\em primordial} $\rho_0$ and $\omega$ mesons. 
In more general terms, the investigation of the event-by-event
fluctuations of particle ratios provides a crucial test of the
hypothetical chemical equilibration -- to see if it also
predicts two particle correlations correctly in addition to 
the single particle inclusive data.

The key point to our argument, that the fluctuation of the $\pi^+/\pi^-$ is
indeed sensitive to the particle numbers
at chemical and {\em not} at thermal 
freeze-out, is the observation \cite{G93,SK97} that the pion number does not change
during the course of the evolution of the system through the hadronic
phase. Typical relaxation times for pion number changing processes are of the
order of 100~fm/c, much longer than the lifetime of the system, which is
about 10~fm/c. In addition, as we shall argue in more detail below, charge
exchange processes, which in principle could affect the
$\pi^+/\pi^-$-fluctuations, lead only to small corrections. 
Finally, considering fluctuations of the multiplicity ratio 
eliminates the effect of volume fluctuations
which are present even with the tightest centrality selection.

Let us now define some notations. 
We define the fluctuation $\delta N_i$ by 
\be N_i = \ave{N_i} + \delta N_i \ee 
where $\ave{N_i}$ is the average number of the particle species $i$.
Then the variance is given by
\begin{equation}
\label{eq:w-def}
\Delta (N_i,N_i) 
\equiv \ave{\delta N_i\delta N_i} 
= \ave{{N_i}^{2}}-\ave{N_i}^{2} 
= w_i \ave{N_i}
\end{equation}
Here we introduced the notation
\begin{equation}
\label{eq:del-def}
\Delta (N_i,N_j)\equiv \ave{N_i\, N_j}-\ave{N_i}\ave{N_j} 
= \ave{\delta N_i \delta N_j}
\end{equation}
and defined $w_i$ to be the ratio $\Delta(N_i,N_i)/\ave{N_i}$. 

The absence of correlations make
the fluctuation in the multiplicity very close to the (inclusive)
average number of particles.  
Hence, in a classical thermal system, \( w_i =1 \), 
since there are no correlations among the particles.  
Bose-Einstein or Fermi-Dirac statistics introduce
correlations so that
\be & w_i =1\pm
\ave{n_i^{2}}/\ave{n_i} &
\\
\inline{with} &\displaystyle 
\ave{n} = \int \frac{d^{3}p}{(2\pi )^{3}}n_{\pm}(p);\, \, \, \, \, \, \, \, 
\ave{n^{2}}=\int \frac{d^{3}p}{(2\pi)^{3}}n^{2}_{\pm }(p) &
\label{eq:thermal_w}
\ee 
where, (\( + \)) stands for Bosons, and (\( - \))
stands for Fermions\cite{Landau5}.  For the systems of interest here, however,
the corrections due to quantum statistics are small; for a pion gas at a
temperature of 170~MeV, \( w_\pi =1.13 \) \cite{Ber94}.

In general however, $w_i$ will {\em not} be equal to
$1\pm\ave{n_i^2}/\ave{n_i}$ due to additional correlations introduced by 
interactions and resonances.  This will be discussed in detail below.

Given the above notation,
the fluctuation of the ratio is given by \cite{BH99}
\be
\label{eq:ratio-fluct}
\frac{\Delta (R_{12},R_{12})}{\ave{R_{12}}^{2}} 
& = &
\left(\frac{\Delta(N_{1},N_{1})}{\ave{N_{1}}^{2}} + 
\frac{\Delta(N_{2},N_{2})}{\ave{N_{2}}^{2}} 
\right.
\non
& & {} \quad
-2
\left. 
\frac{\Delta
    (N_{1},N_{2})}{\ave{N_{1}}\ave{N_{2}}} \right) \ee 
The last
term in Eq.~(\ref{eq:ratio-fluct}) takes into account correlations between the
particles of type 1 and type 2. This term will be important if both particle
types originate from the decay of one and the same resonance. For example, in
case of the $\pi^+/\pi^-$ratio, the $\rho_0$, $\omega$ etc.  
contribute to these correlations. Also, volume fluctuations contribute here. 

\textbf{(i) Volume fluctuations:} Even though data are often selected
according to some centrality trigger, the impact parameter and thus the volume
of the created system, still fluctuates considerably. Assuming, that 
the particle abundance scales linearly with the volume, volume
fluctuations translate directly into fluctuations of the particle number.

However, by considering ratios of particles these fluctuations cancel to
leading order. This can be seen as follows.  Note that we can rewrite 
Eq.~(\ref{eq:ratio-fluct}) as
\be \frac{\Delta (R_{12},R_{12})}{\ave{R_{12}}^{2}} =
\ave{ \left( {\delta N_1 \over \ave{N_1}} - {\delta N_2 \over \ave{N_2}}
  \right)^2 }
\label{eq:ratio_fluct_2}
\ee
Assuming that the volume fluctuation separates from the density
fluctuation, we write 
\be
N 
= n_{\rm ave} (\ave{V} + \delta V)(1 + \delta n/n_{\rm ave})
\ee
where $n_{\rm ave} = \ave{N}/\ave{V}$.
Then to the first order, 
\be
{\delta N \over \ave{N}}
=  
{\delta V\over \ave{V}} + {\delta n \over n_{\rm ave}}
= 
{\delta V\over \ave{V}} + {\delta_n N \over \ave{N}}
\ee
where $\delta_n N = \ave{V}\delta n$ is the number fluctuation due to
the density fluctuation.
Clearly, the volume fluctuation part will be cancelled in 
Eq.~(\ref{eq:ratio_fluct_2}) and hence $\Delta(N_i, N_j)$ in
Eq.~(\ref{eq:ratio-fluct}) can be simply replaced with
\be
\Delta_n(N_i, N_j) = \ave{\delta_n N_i\, \delta_n N_j}
\ee
From now on, unless otherwise signified, 
we will omit the subscript $n$ from $\Delta_n(N_i,N_j)$.

Let us now turn to the discussion of the density fluctuations. 
In the physical system we consider, the density fluctuation is mainly
due to the thermal fluctuation.
As already mentioned above, in absence of any resonances/interactions
the thermal fluctuations 
\( \Delta (N_1, N_1) \) are simply given by Eq.~(\ref{eq:w-def}),
with \( w_1 \) slightly different from unity due to quantum statistics.
Furthermore
in a thermal system the correlation term vanishes, 
i.e. \( \Delta (N_{i,}N_{j})=\delta _{ij}\, \Delta (N_{i},N_{j}) \)
since
$\ave{N_1\,N_2} = \ave{N_1}\ave{N_2}$ in that case.
This changes, however, once interactions,
in particular resonances, are present in the system.

\textbf{(ii) Effect of resonances:} 
A fundamental assumption of the statistical model is that at the
chemical freeze-out time, all the particles including resonances are 
in thermal and chemical equilibrium.  The expansion afterwards
breaks the equilibrium.  However, as discussed above the total number of
$\pi^+$ and $\pi^-$ given by
\begin{equation}
\label{eq:aver-number}
\ave{N_{i}}=
\ave{N_{i}}_T
+
\sum_{R}\ave{R}_T \ave{n_i}_R
\end{equation}
remains constant from this time on.
Here the subscript $T$ on $\langle N_i \rangle_T$ and $\langle R \rangle_T$
denotes the average number of particles and resonances
at the freeze-out time and
$\ave{n_i}_R$ is the average number of the particle type $i$ produced
by the decay of a single resonance $R$.

The presence of resonances which decay
into the particles of interest affects the fluctuations of
each individual particle
(\( \Delta (N_{i},N_{i}) \)) \cite{SRS99}. 
Resonances decaying into both particle species of interest 
also affect the correlation term (\( \Delta (N_{i},N_{j}) \)).
A single resonance contribution 
to \( \Delta (N_{i},N_{j}) \) is given by\cite{JK99_2}
\begin{equation}
\label{eq:resonance-fluct}
{\Delta_R (N_{i},N_{j}) \over \ave{R}}
=
\ave{n_i n_j}_R
+
(w_R - 1)
\ave{n_i}_R
\ave{n_j}_R
\end{equation}
where we defined
\be
\ave{n_i n_j}_R
\equiv
\sum_{r\in {\rm branches}} b^R_r (n^R_i)_r (n^R_j)_r 
\ee
and
\be
\ave{n_i}_R
\equiv
\sum_{r\in {\rm branches}} b^R_r (n^R_i)_r 
\ee
Here the index $r$ runs over all branches,
$b^R_r$ is the branching ratio of the $r$-th branch,
and $(n^R_i)_r$ represents the number of $i$-particles
produced in that decay mode.
At the chemical freeze-out,
the system is in equilibrium and hence
the number of different particle species are uncorrelated.
In the final state where all the the resonances have decayed, the
correlation is given by 
\be
\label{eq:resonance-correlation}
\lefteqn{\Delta(N_i, N_j)
=
w_i^T \ave{N_i}_T \delta_{ij}} &&  
\non
&& {} \quad
+
\sum_{R}
\ave{R}_T
\left(
\ave{n_i n_j}_R
+
(w_R^T - 1)
\ave{n_i}_R
\ave{n_j}_R
\right)
\label{eq:resonance-particle}
\ee
where $w^T_i \langle N_i \rangle_T$ 
denotes the part of the variance due only to the statistical
fluctuations at the chemical freeze-out time.

Putting everything together we get for the fluctuations of the ratio
\begin{equation}
\frac{\Delta (R_{12}, R_{12})}{\ave{R_{12}}^{2}}
=\frac{1}{\ave{N_{2}}}\left( D_{11}+D_{22}-2D_{12}\right) 
\end{equation}
with
\be
D_{11} & = & {\ave{N_2}\over \ave{N_1}}\, F_1
\label{eq:d_11}
\\
D_{22} 
& = & F_2
\\
D_{12} 
& = & 
\sum_R \ave{n_{1} n_{2}}_R {{\ave{R}_T}\over {\ave{N_1}}}
\label{eq:d_+-} 
\ee
where
\be
F_i
& = &
\left( w_{i}^T\, r_{i}
+\sum_{R}\ave{n_{i}^{2}}_R\frac{\ave{R}_T}{\ave{N_i}}\right) 
\non
& = &
\left(1 
+(w_i^T - 1)r_1
+\sum_{R}\left( \ave{n_{i}^{2}}_R - \ave{n_i}_R\right)
\frac{\ave{R}_T}{\ave{N_i}}\right) 
\non
\ee
and we defined $r_i \equiv {\ave{N_i}_T / \ave{N_i}}$.
Here we regarded $(w_R^T - 1)$ to be negligibly small. 
For a typical resonance of
$m_R \sim 1\,{\rm GeV}$, $(w_R^T - 1) < 10^{-2}$ assuming the temperature of
170~MeV. 
(In the numerical results presented below, these terms, though
small, are included.).  
Note that we have factored out $1/ \langle N_2 \rangle$ 
to separate out the explicit dependence on the system size. 

Before we turn to the practical applications, a few comments are in order at
this point.
First,
the effect of the
correlations introduced by the resonances should be most visible
when \( \ave{N_{1}}\simeq \ave{N_{2}} \)
since
the branching fraction \( n_{i}^R \) should
enter with about the same weight.
In this case, a resonance decaying always into a pair of
particles ``1'' and ``2'' contributes about equally  
to $D_{11}$, $D_{22}$ and $D_{12}$ and hence contributes negligibly 
to $(D_{11}{+}D_{22}-2D_{12})$. 
On the other hand, the presence of such resonances does
influence the total number of particle ``2'', $\ave{N_2}$.
Hence, for instance,
$\rho^0$ and $\omega$ will always {\em reduce} the
fluctuation in $\pi^+/\pi^-$ compared to the statistical fluctuation.  
Second,
when \( \ave{N_{2}}\, \gg \, \ave{N_{1}} \),
as in the $K$ to $\pi$ ratio, 
the fluctuation is dominated by the less abundant particle 
type and the resonances feeding into it.

After this general formulation of the problem, let us now turn to the
calculation of the negative to positive pion ratio.
The formalism developed above can be easily applied to the case of 
\( {\pi ^{+}/\pi ^{-}} \)fluctuations
by setting
\[
N_{1}=\pi ^{+},\, \, \, \, \, N_{2}=\pi ^{-}\]
Typically, \( \displaystyle {\ave{\pi^{+}}}/{\ave{\pi^{-}}}\simeq 1 \).
In table (\ref{tab:contrib}) we show the most important 
contributions from hadronic resonances. 
This calculation includes mesons and baryons up to $m \sim 1.5$~GeV
as listed in the particle data book.
Weakly decaying strange particles are regarded as stable, 
but letting them decay changes the main result very little.
The values of temperature, baryon chemical potential and the strangeness
chemical potential are the same as those in \cite{BHS99}, i.e.,
$T=170$~MeV, $\mu_b=270$~MeV and $\mu_s=74$~MeV.

As already pointed out, the \( \rho ^{0} \) and \( \omega \) contribute
about 50 \% of the correlations.  Furthermore, the correlation-term 
$2D_{+-}$ is
\emph{seventy percent} 
of the individual contributions $D_{++}$ or $D_{--}$.
This is a
sizable correction which should be visible in experiment.  Furthermore,
since only very few resonances have decay channels with more than one charged
pion of the the same charge, $\ave{n_\pm^2}_R - \ave{n_\pm}_R \simeq 0$ to a
good approximation.  
Also, $(w_\pm^T - 1)r_\pm \simeq 4\ \%$ at $T = 170\ {\rm MeV}$.  
Hence, the fluctuations of the individual pion contributions are
very close to the statistical limit of \( D_{++} \simeq D_{--} \simeq 1. \)
Thus the resonances contribute predominantly to the correlation term $D_{+-}$.
\begin{minipage}{8.5cm}
\begin{table}[htb]
\caption{Contributions from different hadrons to the fluctuations of the
  $\pi^+ / \pi^-$ ratio. Contributions are given in fraction of the 
total result. 
\label{tab:contrib}}
\vspace{3mm}
{\centering \begin{tabular}{|c|c|c|c|c|c|}
Particle&
\( \frac{D_{++}}{D_{++}^{tot}} \)&
\( \frac{D_{--}}{D_{--}^{tot}} \)&
\( \frac{D_{+-}}{D_{+-}^{tot}} \)&
\( \frac{n_{+}}{n_{+}^{tot}} \)&
\( \frac{n_{-}}{n_{-}^{tot}} \)
\\
\hline 
\hline 
\( \pi^+  \)&
0.32 &   0.00 &   0.00 &   0.31 &   0.00
\\
\hline 
\( \pi^-  \)&
0.00 &   0.32 &   0.00 &   0.00 &   0.31
\\
\hline 
\( \eta \)&
0.02 &   0.02 &   0.06 &   0.02 &   0.02
\\
\hline 
\( \rho^+  \)&
0.08 &   0.00 &   0.00 &   0.09 &   0.00
\\
\hline 
\( \rho^0  \)&
0.08 &   0.08 &   0.24 &   0.09 &   0.09
\\
\hline 
\( \rho^-  \)&
0.00 &   0.08 &   0.00 &   0.00 &   0.09
\\
\hline 
\( \omega  \)&
0.07 &   0.07 &   0.21 &   0.08 &   0.08
\\
\hline 
Others &
0.44 & 0.44 & 0.55 & 0.43 & 0.44
\\
\end{tabular}}
\end{table}
\end{minipage}
\begin{minipage}{8.5cm}
\begin{table}[htb]
\caption{
Total values}
\vspace{3mm}
{\centering \begin{tabular}{|c|c|c|c|c|c|}
&
\( {D_{++}^{tot}} \)&
\( {D_{--}^{tot}} \)&
\( {2D_{+-}^{tot}} \)&
\( {n_{+}^{tot}} \) &
\( {n_{-}^{tot}} \) 
\\
\hline 
\hline 
Values
&
1.09 & 1.09 & 0.76 & $0.20\,{\rm fm}^{-3} $ & $0.20\,{\rm fm}^{-3}$
\\
\end{tabular}}
\end{table}
\end{minipage}

Using $\ave{\pi^+} = \ave{\pi^-} = \ave{\pi^+ + \pi^-}/2$ and 
$\ave{\pi^+ +  \pi^-} = 220$ from the recent NA49 results \cite{NA49}
on event-by-event
fluctuations of the transverse momentum and the $K/\pi$-ratio, we obtain for 
the total fluctuation $\pi^+/\pi^-$ ratio

\be
\frac{\Delta (R_{+-},R_{+-})}{\ave{R_{+-}}^{2}}\equiv
\sigma _{+-}^{2}= 0.0128 
\ee
so that $\sigma _{+-}= 0.113$
A more useful quantity, to consider, however, is the ratio of the above total
fluctuations over the purely statistical value. The latter can be possibly
obtained in experiment by the analysis of mixed events.
Possible artificial contributions due to
experimental uncertainties, such as particle identification etc. should cancel
to a large extent in this ratio, and thus a comparison with theory becomes
more meaningful. 
The value for the statistical fluctuation for us is simply given by
\be
\left(\sigma_{+-}^{stat.}\right)^2 
= \frac{1}{\ave{\pi^+}} + \frac{1}{\ave{\pi^-}} 
\ee 
since in mixed events all correlations, even the quantum statistical ones, are
absent, i.e. $w_+ = w_- = 1$.
Our result for this ratio is 
\be
{\sigma_{+-}^2 \over ( \sigma_{+-}^{stat} )^2 } = 0.70
\ee
Note that this ratio is independent of $\ave{\pi^{-}}$.
Thus the correlations introduced by the presence of resonances {\em reduce} 
the fluctuations by 30 \%. Looking at
table (\ref{tab:contrib}), the most important contribution of the correlation
come from the $\rho$ and $\omega$ meson. Or in other words, the measurement of
the fluctuations of the $\pi^+/\pi^-$ provides a strong constraint on the
initial number of $\rho^0$ and $\omega$-mesons.   

Doing the same analysis for the $K/\pi$ ratio, where preliminary data exist
\cite{NA49}, we find
\be
\frac{\sigma_{K/\pi}^2}{(\sigma_{K/\pi}^{stat})^2} = 1.04
\ee
in good agreement with the data.
Note, however that our value for the fluctuation itself, 
$\sigma_{K/\pi} = 0.17$ differs from the experimental value of
$\sigma^{exp}_{K/\pi} = 0.23$ indicating the effect of additional
fluctuations from particle identifications etc.\cite{Roland}.
The reason (see \cite{JK99_2} for
details), is that in case of the $K/\pi$ ratio, the fluctuation is
dominated by the contribution from the kaon, which is largest due to the
small abundance.

Let us close by discussing some possible caveats. First, there is the question
of acceptance cuts. Clearly, one should {\em not} consider the particle ratio
of the full $4 \pi$ acceptance. In this case charge conservation will impose
severe constraints and reduce the fluctuations of the $\pi^+ / \pi^-$
ratio. However, as long as a limited acceptance in, say, rapidity is
considered, the constraint from charge conservation is minimal and our
assumption of a grand-canonical ensemble are well justified. Limited
acceptance on the other hand may reduce the effect of resonances on the
correlation term, as some of the decay products may end up outside the
acceptance. In order to estimate this effect we have performed a Monte-Carlo
study. Given the rapidity distribution
of charged particles for a Pb+Pb collision at SPS-energies \cite{NA49}, 
a rapidity window
of $\Delta y = 1$ changes the above results by less then one percent. On the
other hand this window covers only a fraction of the observed rapidity
distribution, so that constraints from charge conservation are negligible.
Finally there are the charge exchange reactions such as $\pi^+{+}\pi^-
\leftrightarrow \pi^0{+}\pi^0$. These reactions in principle could change the  
$\pi^+ / \pi^-$-ratio in a given event. However, detailed balance requires the
net change to be close to zero. In addition,
for a given event, these reactions influence {\em not} the difference between 
$\pi^+$ and $\pi^-$ but only their sum.
Thus, again we expect only small corrections to the above
result since the sum is large compared to the expected changes. 

The reaction with baryons, notably 
$\pi^-{+}p \leftrightarrow \pi^0{+}n$ might be more effective. But again,
detailed balance and the fact that the nucleon/pion ratios is so small should
make these corrections insignificant. A detailed quantitative investigation of
these corrections will be presented in \cite{JK99_2}.

In conclusion, we propose to study the event-by-event fluctuations of the
$\pi^+/\pi^-$-ratio in relativistic heavy ion collisions. This measurement will
provide important information about the abundance of short lived resonances
right after hadronization and/or chemical freeze-out. It will further impose a
strong test on the validity of the chemical equilibration hypothesis in
these reactions. If chemical equilibrium is reached with the values for
temperature and chemical potential extracted from the single particle
distributions, we predict that the fluctuations of the  $\pi^+/\pi^-$-ratio
should be about 70~\% of the statistical ones. 
It would be also of great interest to study these fluctuations in proton-proton
and peripheral heavy ion collisions, where the particle abundances also seem
to indicate chemical equilibrium.  

In addition, any value of the 
$\sigma_{+-}/\sigma_{+-}^{stat}$ significantly larger than 1 cannot be 
explained with a simple hadronic gas picture and thus would indicate 
new physics.  
One possible scenario might be the Quark Gluon Plasma bubble
formation \cite{BH99}.
Thus the $\pi^+/\pi^-$ ratio as a function of $E_T$ may serve
as an alternative signal for the QGP.

We would like to thank G. Roland and S. Voloshin for useful comments.  
This work was supported by the Director, 
Office of Energy Research, Office of High Energy and Nuclear Physics, 
Division of Nuclear Physics, and by the Office of Basic Energy
Sciences, Division of Nuclear Sciences, of the U.S. Department of Energy 
under Contract No. DE-AC03-76SF00098.


\end{multicols}

\end{document}